\documentclass[journal=jacsat,manuscript=article]{achemso}

\usepackage[version=3]{mhchem} 

\usepackage{graphicx}
\usepackage{dcolumn}
\usepackage{bm}

\usepackage[monochrome]{xcolor} 

\usepackage[utf8]{inputenc}
\usepackage[T1]{fontenc}
\usepackage{mathptmx}
\usepackage{etoolbox}
\mathchardef\mhy="2D

\author{Randy Sterbentz}
\author{Bogyeom Kim}
\author{Anayeli Flores-Garibay}
\author{Kristine L. Haley}
\author{Nicholas T. Pereira}
\affiliation{Department of Physics and Astronomy, University of Nevada Las Vegas, Las Vegas, NV 89154, USA}
\author{Kenji Watanabe}
\affiliation{
Research Center for Electronic and Optical Materials, National Institute for Materials Science, 1-1 Namiki, Tsukuba 305-0044, Japan}
\author{Takashi Taniguchi}
\affiliation{
Research Center for Materials Nanoarchitectonics, National Institute for Materials Science,  1-1 Namiki, Tsukuba 305-0044, Japan}
\author{Joshua O. Island}
\email{joshua.island@unlv.edu}
\affiliation{Department of Physics and Astronomy, University of Nevada Las Vegas, Las Vegas, NV 89154, USA}

\title[Gating \textcolor{red}{monolayer and bilayer graphene} with a two-dimensional semiconductor]{Gating \textcolor{red}{monolayer and bilayer graphene} with \textcolor{red}{a two-dimensional semiconductor}}

\keywords{Graphene, MoS$_2$, MoSe$_2$, WS$_2$, WSe$_2$, TMD, gating, semiconductor, voltage clamp}

\begin{document}








\begin{abstract}
Metals are commonly used as electrostatic gates in devices due to their abundant charge carrier densities that are necessary for efficient charging and discharging. A semiconducting gate can be beneficial for certain fabrication processes, in low light conditions, and for specific gating properties. \textcolor{red}{We determine the effectiveness and limitations of a semiconducting gate in graphene and bilayer graphene devices. Using the semiconducting transition metal dichalcogenides molybdenum disulfide (MoS$_2$), molybdenum diselenide (MoSe$_2$), tungsten disulfide (WS$_2$), and tungsten diselenide (WSe$_2$), we show that two-dimensional semiconductors can be used to suitably gate the graphene devices under appropriate operating conditions. For single-gated devices, semiconducting gates are comparable to metallic gates below liquid helium temperatures but include resistivity features resulting from gate voltage clamping of the semiconductor. In dual-gated devices, we pin down the parameter range of effective operation and find that the semiconducting depletion regime results in clamping and hysteresis from defect-state charge trapping.}
\end{abstract}

\maketitle

\section{Introduction}\label{intro}
Electrostatic gating in graphene devices has evolved over the years. The earliest devices employed highly doped silicon substrates as global back gates or evaporated metals for local top and bottom gates\cite{novoselov2004, zhang2005experimental, zhang2009}. Recently, few-layer graphene has emerged as an ideal gate material for its atomically flat and chemically inert character. This leads to an overall reduction in charge inhomogeneity and observation of intricate correlated states when directly compared with deposited metal gates\cite{zibrov2017tunable}. 
There are situations though in which a semiconducting gate may be more appropriate. Semiconducting gates employed in high mobility transistors provide over voltage protection and lower leakage current when compared with metallic gates\cite{qian2019}. In novel high-frequency, on-chip terahertz measurements of graphene, a semiconducting gate is used to avoid absorption of the probing field\cite{gallagher2019, seo2024on-chip}. Semiconducting gates could also play an important role in sensitive photodetectors and bolometers requiring minimal absorption or reflection from a top gate electrode\cite{yan2012dual, yoshioka2022ultrafast}. They could boost the already impressive figures of merit for dual gated bilayer graphene bolometers that take advantage of intrinsic bulk response by minimizing top gate interference thereby further increasing sensitivity\cite{yan2012dual}. The increased resistance of a semiconducting top gate would also be advantageous in high-bandwidth graphene photodetectors by reducing the RC time constant derived from the gate capacitance\cite{yoshioka2022ultrafast}. Furthermore, in comparison with semi-transparent nichrome or zinc oxide gates that are grown or deposited\cite{kim2013photothermal, gopalan2017mid}, 2D material stacks provide sharp, charge homogeneous interfaces. These examples highlight the novelty of using a semiconducting gate but it has not been shown, given the depletion characteristics of a semiconductor, for what conditions electrostatic gating should function properly in a graphene device. 

In a simple parallel-plate capacitor configuration with a metal and a two-dimensional electron gas (like graphene), applying a gate voltage ($V_G$) results in a change in the chemical potential ($\mu$) in the graphene and the electrostatic potential ($\phi$) between the two layers: $eV_G=e\phi+\mu$, where $\phi$ is determined by the geometric capacitance, $\phi=ne/C_G$. If a semiconductor replaces the metal as a gate, the reduction in overall charge density and presence of an electronic band gap will alter the gate response. As the Fermi level is driven into the band gap of the semiconductor while sweeping the gate contact voltage, the effective potential between the gate and the graphene channel will remain unchanged due to a lack of charge carriers in the gate itself. This voltage clamping effect of the gate will have meaningful consequences on the transport characteristics of a graphene device. Moreover, the temperature of the system becomes relevant as the threshold voltage in the gate changes and the thermal energy of carriers varies. This implies that the effect of sweeping beyond the threshold voltage of the semiconducting gate should be more pronounced at lower temperatures.  

Here we investigate the viability of using \textcolor{red}{2D transition metal dichalcogenides (MoS$_2$. MoSe$_2$, WS$_2$, WSe$_2$)} as an electrostatic gate for monolayer and bilayer graphene (BLG) devices. MoS$_2$ \textcolor{red}{is highlighted in the main text} because it has the smallest difference in its electron affinity and graphene's work function when compared with the other transition metal dichalcogenides\cite{liu2016van, Cloninger_2024}. This results in \textcolor{red}{ON state characteristics} closest to zero gate voltage. In a single-gate response, we show that both mono- and bilayer graphene can be effectively gated down to liquid helium temperatures, albeit with resistance artifacts associated with clamping of the semiconducting gate. Dual-gated bilayer graphene with a few-layer graphene (FLG) control gate, allows us to clearly demarcate the parameter space applicable to efficient gating. A 1D potential model is modified to support our experimental results and corroborate gate voltages at which clamping occurs. Additionally, we observe significant hysteresis for clamped gate voltages that can be attributed to trap states from intrinsic defects. The onset of hysteresis is directly correlated with the threshold voltage of the semiconducting gate. \textcolor{red}{Finally, we show that all four materials can be used to gate bilayer graphene and we compare their differences.} Our results provide specific guidelines for gating using semiconducting gates in graphene devices and pave the way for their use in \textcolor{red}{sensitive detectors and spectroscopy.} 

\section{Results}
\subsection{Single-gate characteristics}

We construct 2D van der Waals heterostructures to assess the feasibility of a semiconducting gate. \textcolor{red}{The devices are created using a dry stacking and transfer technique\cite{purdie2018cleaning, haley2021heated} and a summary of the seven devices measured in this study is shown in Supplementary Table 1 and Supplementary Figure 1.} \textcolor{red}{Flakes of hexagonal boron nitride (h-BN) are employed as dielectric layers separating the gates from the graphene layers for all devices investigated. The Methods section presents the fabrication and measurement details for all devices studied.} The single- and dual-gate results for bilayer graphene are presented here in the main text and results for a similar monolayer graphene device \textcolor{red}{with an MoS$_2$ gate} can be found in Supplementary Figure 2. Figure \ref{fig:1}(a) shows an optical image of a bilayer device \textcolor{red}{with an MoS$_2$ gate} with its stacking order shown as a model in the inset. The resistivity is measured using either the MoS$_2$ top gate (red layer in Figure \ref{fig:1}(a)) or the few-layer graphene (FLG) control gate (blue layer in Figure \ref{fig:1}(a)). By utilizing both the FLG and MoS$_2$ gates in a single device, we can directly compare the well-known gating behavior of FLG with the unknown gating characteristics of the semiconductor.

Figure \ref{fig:1}(b) shows the individual gate-dependent transport characteristics of BLG at room temperature. Both the MoS$_2$ gate and FLG gate, when swept separately, yield the characteristic peak in resistivity at the charge neutrality point (CNP)\cite{geim2007, allen2010, yan2010, kuzmenko2009, efetov2011}. The resistivity reported serves as an upper bound for the device as it includes the ungated regions of the BLG channel as well. The FLG gate achieves a lower resistivity at higher doping than the MoS$_2$ gate due to a larger area of overlap between the BLG and FLG flakes (the overlap can be seen in Figure \ref{fig:1}(a)). While the FLG gate shows no significant hysteresis, the MoS$_2$ gate response displays a shoulder feature with hysteresis between forward (solid) and backward (dashed) sweeps. These are associated with voltage clamping in the MoS$_2$ and explained in more detail below. Figure \ref{fig:1}(c-d) present color maps of the gate-dependent transport at temperatures from 400 K down to 2.5 K. The CNP is evident as a central peak and does not shift as temperature decreases for either gate, highlighting the capability of MoS$_2$ for gating down to liquid helium temperatures. The CNP is situated to the left of zero gate voltage indicating some residual doping noticeable for both gates. From the CNP position, we determine the device has a residual n-doping of 0.126$\times$10$^{12}$ cm$^{-2}$ possibly due to trapped contaminants at the BLG layer.

We extract the field-effect mobilities for the BLG holes and electrons from these gate sweeps to better characterize and compare the effects of both gates. Mobilities are calculated based on the Drude model\cite{gosling2021} by taking the steepest slope of conductivity versus carrier density and using $\mu_{(h,e)} = \frac{1}{e}\frac{\partial\sigma}{\partial n_{(h,e)}}$. Figure \ref{fig:1}(e-f) show the extracted mobilities as a function of temperature. Throughout the temperature range tested, the MoS$_2$ gate yields mobilities comparable to the FLG gate, with values similar to previous studies\cite{cobaleda2014, tan2022,shimazaki2015}. This confirms the efficacy of MoS$_2$ to operate as a gate over a wide range of temperatures.

\subsection{Dual-gate characteristics}

To more thoroughly investigate the parameter space, we expand our measurement to varying both gates simultaneously. The results of these 2D gate sweeps are presented in Figure \ref{fig:2}, where panels (a-e) show the MoS$_2$ as the fast forward sweep axis (from negative to positive voltage), and panels (f-j) show it as the fast backward sweep axis. At 400 K (a,f), the BLG resistivity shows two ridges of high resistivity. These ridges correspond to CNPs of different regions of the BLG: the horizontal line about $V_{FLG}=0$ V corresponds to the region singly-gated by the FLG gate, and the diagonal line corresponds to the dual-gated region. We expect the CNP of the dual-gated region to occur where electrostatic doping by one gate is canceled by the other gate\cite{shimazaki2015, yan2010, henriksen2010}, creating a diagonal line with negative slope across the 2D plot. The total carrier concentration $n$ in the BLG is calculated from the sum of the individual gate influences: $n = (C'_{MoS_2}V_{MoS_2} + C'_{FLG}V_{FLG})/e$, where $C'_i$ is the capacitance per unit area between the BLG and gate $i$, and $V_i$ is the voltage applied to gate $i$  ($i\in\{MoS_2,FLG\}$). The slope of the CNP line (where $n = 0$ cm$^{-2}$) is $-C'_{MoS_2}/C'_{FLG} = -d_{FLG}/d_{MoS_2}$, where $d_i$ is the thickness of dielectric layer between the graphene and gate $i$. An apparent linear portion of the CNP exists in quadrant II of Figure \ref{fig:2}(a). Fitting to a line results in a slope of $-1.44$. This value closely matches the h-BN thickness ratio measured by AFM of $-d_{FLG}/d_{MoS_2} = -1.34$. 

The definition of $n$ assumes the charge density in each gate varies linearly with applied voltage, but this is not true for the semiconducting MoS$_2$ gate. This can be seen in quadrant IV of the 2D gate sweeps where the CNP becomes disjoint from that in quadrant II. The shift suggests a limitation of the MoS$_2$ gate. As the temperature decreases, the shift appears larger, especially in the forward sweep (Figure \ref{fig:2}(b-e)). Other features appear at lower temperatures as well, such as CNP splitting and non-monotonic CNP contours. These features are reproducible, but depend on both the sweep direction and which gate is being swept fast (MoS$_2$ or FLG) (see Supplementary Figure 3). They coincide with the features seen in the 1D gate sweeps of Figure \ref{fig:1}, thus indicating single-gate sweeps must be carefully examined to assess limitations of the MoS$_2$ gate. 

We perform further analysis of the 2D gate sweeps by looking at the hysteresis of the CNP feature with respect to the MoS$_2$ gate voltage. Figure \ref{fig:2}(k,l) show the hysteresis $\Delta V_{MoS_2,\textrm{CNP}}$ as a function of temperature (k) and as a function of the FLG gate voltage (l). From panel (k), we see the magnitude of the hysteresis increases as temperature is decreased. Panel (l) shows for $V_{FLG}>0$ V (corresponding to quadrant II of the 2D gate sweeps), hysteresis is minimal, but sharply increases at lower gate voltages. Notably, the hysteresis appears to drop back to zero at large negative FLG voltages, suggesting that gating becomes effective again. However, in Figure \ref{fig:2}(m), we see the new CNP settles to a gate voltage that corresponds to a p-doped value of $\sim2\times10^{12}$ cm$^{-2}$. 

Within the same device, we can measure the transport characteristics of the MoS$_2$ to determine properties such as its threshold voltage and its sheet conductivity\cite{late2012, cho2013, lee2015}. For these measurements, two graphite contacts to the MoS$_2$ layer are used. Figure \ref{fig:3}(a) shows the transconductance of the MoS$_2$ layer as a function of voltage applied to the BLG and FLG tied together, using both of them as a single global back gate. The conductivity measurements have a noise floor of approximately 10 pS. Most of the transconductance scans fall below this value in the MoS$_2$ off-state. We determine an on-off ratio of at least 10$^6$ and a field-effect mobility of roughly 30-70 cm$^2$/Vs depending on temperature, which is comparable to other reports\cite{late2012, jariwala2013, dibartolomeo2018, lin2016, radisavljevic2013, yang2020, ji2019}. At temperatures above 360 K, a notable leakage current dominates the off-state signal (see Supplementary Figure 4). The inset of Figure \ref{fig:3}(a) shows the threshold gate voltage versus temperature, extracted by fitting the linear region of the transconductance with a line and finding its $x$-intercept\cite{late2012, lee2015}. The generally monotonic trend aligns with the expectation that thermally excited electrons contribute to turning on MoS$_2$ at lower gate voltages\cite{ji2019, radisavljevic2013}.

\subsection{Semiconductor characteristics}
To better understand the limitations of the MoS$_2$ gate during electrical measurements of the BLG, we examine the MoS$_2$ characteristics for the same gating conditions shown in Figure \ref{fig:2}. Figure \ref{fig:3}(b-c) show the conductivity of MoS$_2$ as a function of gate voltages relative to the BLG at 2.5 K and 300 K. This was achieved by setting a constant bias voltage $V_{DS}$ across the MoS$_2$ while varying the potential on the BLG and FLG flakes. The relative potentials were adjusted to match the gate voltages applied during the 2D gate sweeps of the BLG resistivity. The voltages can be converted as follows:
\begin{equation}
    V_{MoS_2\mhy BLG} = -V_{BLG\mhy MoS_2}
\end{equation}
\begin{equation}
    V_{FLG\mhy BLG} = V_{FLG\mhy MoS_2} - V_{BLG\mhy MoS_2}
\end{equation}
where the convention $V_{A\mhy B}$ indicates a potential difference from layer $A$ (positive side) to layer $B$ (negative side). The dashed lines of Figure \ref{fig:3}(b-c) indicate the MoS$_2$ threshold voltage, determined by extracting the $V_{MoS_2\mhy BLG}$ corresponding to a MoS$_2$ conductivity just above the noise floor. This line effectively demarcates the transition between the on-state and off-state of the MoS$_2$ for the same gating conditions in Figure 2.

We match the on-off state threshold voltage of the MoS$_2$ with the hysteresis of the BLG resistivity and see a distinct overlap. Figure \ref{fig:3}(d-e) shows the difference in BLG resistivity between the forward sweep and backward sweep at 2.5 K and 300 K with the threshold voltage determined from panels (b-c) superimposed. Qualitatively, there is a distinct change in behavior of the CNP hysteresis when MoS$_2$ is in its on-state versus its off-state. Further analysis of low temperature hysteresis dependence on gate voltage sweep rate is available in Supplementary Figure 5. In the off-state the hysteresis varies greatly. The forward and backward sweeps significantly misalign, as indicated by the divergence between the positive and negative values in the plot.

We highlight the success of the on-state MoS$_2$ gate by measuring the band gap opening of BLG as a function of displacement field, as shown in Supplementary Figure 6. In the on-state of the MoS$_2$ gate, we determine a band gap of 53 meV at a displacement field magnitude of 0.77 V/nm, which is comparable to previously reported values measured with metallic gates\cite{zhang2009}. The onset of hysteresis in the BLG can be explained by a voltage clamping effect of the MoS$_2$ gate. A 1D model calculating the effective potential between the BLG channel and the MoS$_2$ gate for different gate contact voltages has been adapted from Qian et al.\cite{qian2019} and is discussed in Supplementary Note 7, see also Supplementary Figure 7. We find in the MoS$_2$ on-state, the potential follows the applied gate contact voltage as expected. However, upon entering the off-state above a threshold voltage of 1.2 V, the potential becomes clamped and remains constant for any applied gate contact voltage. This leads to the unchanged charge neutrality point in the forward sweeps shown in Figures 2 and 3. The hysteresis observed between forward and backward sweeps may be due to intrinsic defects in the MoS$_2$ that act as trap states. The hysteresis reaches a maximum within the band gap of the MoS$_2$ and corresponds to a trap state density of $\approx 2\times10^{12}$ cm$^{-2}$ (Supplementary Figure 4(d)). This agrees with the reported magnitude of sulfur defect densities in MoS$_2$ flakes at $2.9\times10^{12}$ cm$^{-2}$\cite{trainer2022visualization}. At the highest positive MoS$_2$ contact gate voltages and for finite negative FLG gate voltages, we observe an unclamping of the MoS$_2$ gate and re-emergence of a hysteresis-free CNP (Figure \ref{fig:2}(l) and Figure \ref{fig:3}(e)). We conjecture that the Fermi level in the MoS$_2$ gate reaches the valence band as a result of partial gating from the FLG flake through the low density BLG layer. This occurs due to incomplete screening from a sandwiched 2D metal\cite{luryi1988quantum}. The splitting of the CNP at the lowest temperatures explored (Figures \ref{fig:2}(e,j) and Figure \ref{fig:3}(d)) is not expected but may be due to a crack in the MoS$_2$ flake causing two slightly different gating responses. We have observed cracking of various transition metal dichalcogenides during the van der Waals heterostructure fabrication process.  

\subsection{Comparison with MoSe$_2$, WS$_2$, and WSe$_2$}
\textcolor{red}{For direct comparison, we also present graphene devices utilizing MoSe$_2$, WS$_2$, and WSe$_2$ as semiconductor gates. The room temperature characteristics of three bilayer devices are presented in Figure \ref{fig:4} and two additional MoSe$_2$ devices, as well as low temperature data, can be found in Supplementary Figure 8. The architecture for these devices is the same as the MoS$_2$ bilayer device, with a FLG bottom gate used as a control gate. Figure \ref{fig:4}(a) compares the transistor characteristics at room temperature of the four devices using four different materials. The MoS$_2$ curve is replotted from Figure \ref{fig:3}(a). MoS$_2$ and MoSe$_2$ present n-type carrier response, WSe$_2$ presents p-type response, and WS$_2$ presents ambipolar response. In Figure \ref{fig:4}(b-d), the BLG resistivity is plotted as a function of the semiconducting gate and the FLG control gate. As in Figure 2, a white dashed line is plotted in Figure \ref{fig:4}(b-d) to identify the expected position of the CNP of BLG given two metallic gates. In the case of MSe$_2$, the CNP becomes disjoint for electron transport in the BLG, similar to the MoS$_2$ device, while for WSe$_2$, the CNP becomes disjoint for hole transport. Given our analysis of the MoS$_2$ device, these are expected behaviors due to the n- and p-type response of these two semiconductors. The WS$_2$ device displays a relatively smaller shift in the CNP compared with the other materials due to its ambipolar response. 
The corresponding hysteresis in forward and backward gate sweeps are plotted in Figure \ref{fig:4}(e-g) for MoSe$_2$, WS$_2$, and WSe$_2$, respectively. The limits of the colorscale have been chosen to match those in Figure \ref{fig:3}(e) for a better comparison between the four materials but note that the intensity of the subtracted hysteresis will also depend on the BLG resistivity itself. In comparison with the MoS$_2$ device which had minimal hysteresis in the ON state at room temperature, these new materials present a greater range of hysteresis that spans the gate voltage parameter space. }

\section{Discussion}
\textcolor{red}{The comparison of all four materials invites strategic engineering of novel devices. For example, if electron transport and investigation around charge neutrality is primarily desired in a graphene device with a semiconducting gate, MoS$_2$ is the likely candidate, as it provides access to the largest range of density change with hysteresis-free gating. If, on the other hand, p-type transport is desired, WSe$_2$ provides the best characteristics. WS$_2$ presents a unique choice if amibpolar transport is needed away from charge neutrality. These general considerations are guidelines for device architecture, but details about the semiconducting gates must be considered. Our semiconducting gates have a range of thicknesses from bilayer to bulk-like (>6 layers). They are also contacted using few-layer graphene flakes, which determines the Schottky barrier between the semiconducting gate and its contact electrode. If different contact materials are used or the semiconducting gate enters the few-layer regime where the band gap is known to change\cite{mak2010atomically, splendiani2010emerging}, the detailed gating characteristics of a graphene device will also likely change. }

We have demonstrated the ability of \textcolor{red}{semiconducting transition metal dichalcogenides} to electrostatically gate mono- and bilayer graphene. As a single gate, MoS$_2$ can effectively induce doping in graphene and bilayer graphene down to temperatures as low as 2.5 K. A shoulder feature in the 1D gate sweeps is identified and associated with voltage clamping in the MoS$_2$ gate. We find similar mobilities when using either the MoS$_2$ gate or a FLG control gate at high doping. In a dual-gate configuration, we identify a region of gate voltage parameter space where MoS$_2$ is in its on-state and provides adequate gating response. In the off-state, we observe voltage clamping of the MoS$_2$ gate leading to an unexpected deviation of the CNP from the linear trend in 2D gate parameter space. A 1D potential model corroborates this clamping effect. Hysteresis of the CNP in the MoS$_2$ off-state agrees with filling and emptying of trap states from intrinsic sulfur vacancies. \textcolor{red}{Finally, we directly compare bilayer graphene devices using other transition metal dichalcogenides (MoSe$_2$, WS$_2$, and WSe$_2$) showing unqiue n-type, p-type, and ambipolar characteristics. The advantages and disadvantages of using each material is discussed.} Our results provide straightforward guidelines for the operation of a semiconducting gate in graphene devices and pave the way toward novel optoelectronic device architectures, high-frequency on-chip measurements, and sensitive detectors. 

\section{Methods}
Materials were isolated from bulk crystals via mechanical exfoliation\cite{novoselov2004}. Exfoliated flakes were then assembled into a vertical heterostructure using a dry stacking method\cite{purdie2018cleaning, haley2021}. Few-layer graphene (graphite) flakes were used as contacts to the semiconducting gates to allow independent measurement of them. The final stack was fully encapsulated with h-BN. h-BN thicknesses are chosen according to optical contrast\cite{golla2013optical} and range from 20 to 70 nm. For the main text MoS$_2$ device, the thicknesses of the h-BN were measured using atomic force microscopy for use in the 1D potential model ($d_{MoS_2}$ = 26.7 nm and $d_{FLG}$ = 35.9 nm). A dielectric constant of $\epsilon_{BN}$ = 3.4$\epsilon_0$ was also used, where $\epsilon_0$ is the permittivity of free space\cite{pierret2022}. A summary of the devices measured is presented in Supplementary Table 1 and optical microscopy images are shown in Supplementary Figure 1. 
Standard photolithographic methods were used to pattern the electrodes. Graphite/graphene layers were exposed by etching away the h-BN in a CHF$_3$/O$_2$ plasma, then metal electrodes were either sputtered (Au 50 nm) at 5 mTorr or deposited by electron-beam evaporation (Cr 5 nm, Au 45 nm) at 10$^{-6}$ Torr. Measurements were performed in a Quantum Design Evercool II PPMS and a Quantum Design Opticool cryostat, capable of base temperatures $\sim$2.0 K. BLG resistivity measurements were performed in a two (TMDG026) or four-probe (all others) configuration. Semiconducting gate conductivity measurements were performed in a two-probe configuration.

\section{Data Availability Statement}
The data that support the findings of this study are available from the corresponding author upon reasonable request.

\section{Acknowledgements}
This work was supported by the National Science Foundation under Grant No. (2047509) and by, or in part by, the U.S. Army Research Laboratory and the U.S. Army Research Office under contract/grant number (W911NF2310160). K.W. and T.T. acknowledge support from the JSPS KAKENHI (Grant Numbers 21H05233 and 23H02052) and World Premier International Research Center Initiative (WPI), MEXT, Japan.

\section{Author Contribution}
R.S., B.K., A.F., K.L.H., and N.T.P. fabricated the devices. R.S. performed the measurements and modified the 1D gating model. K.W. and T.T. grew the hBN crystals. R.S. and J.O.I. wrote the manuscript in consultation with K.L.H. and N.T.P.

\section{Competing interests}
The authors declare no competing interests.

\bibliography{main_text}

\providecommand{\noopsort}[1]{}\providecommand{\singleletter}[1]{#1}%
\providecommand{\latin}[1]{#1}
\makeatletter
\providecommand{\doi}
  {\begingroup\let\do\@makeother\dospecials
  \catcode`\{=1 \catcode`\}=2 \doi@aux}
\providecommand{\doi@aux}[1]{\endgroup\texttt{#1}}
\makeatother
\providecommand*\mcitethebibliography{\thebibliography}
\csname @ifundefined\endcsname{endmcitethebibliography}  {\let\endmcitethebibliography\endthebibliography}{}
\begin{mcitethebibliography}{42}
\providecommand*\natexlab[1]{#1}
\providecommand*\mciteSetBstSublistMode[1]{}
\providecommand*\mciteSetBstMaxWidthForm[2]{}
\providecommand*\mciteBstWouldAddEndPuncttrue
  {\def\EndOfBibitem{\unskip.}}
\providecommand*\mciteBstWouldAddEndPunctfalse
  {\let\EndOfBibitem\relax}
\providecommand*\mciteSetBstMidEndSepPunct[3]{}
\providecommand*\mciteSetBstSublistLabelBeginEnd[3]{}
\providecommand*\EndOfBibitem{}
\mciteSetBstSublistMode{f}
\mciteSetBstMaxWidthForm{subitem}{(\alph{mcitesubitemcount})}
\mciteSetBstSublistLabelBeginEnd
  {\mcitemaxwidthsubitemform\space}
  {\relax}
  {\relax}

\bibitem[Novoselov \latin{et~al.}(2004)Novoselov, Geim, Morozov, Jiang, Zhang, Dubonos, Grigorieva, and Firsov]{novoselov2004}
Novoselov,~K.~S.; Geim,~A.~K.; Morozov,~S.~V.; Jiang,~D.; Zhang,~Y.; Dubonos,~S.~V.; Grigorieva,~I.~V.; Firsov,~A.~A. Electric Field Effect in Atomically Thin Carbon Films. \emph{Science} \textbf{2004}, \emph{306}, 666--669\relax
\mciteBstWouldAddEndPuncttrue
\mciteSetBstMidEndSepPunct{\mcitedefaultmidpunct}
{\mcitedefaultendpunct}{\mcitedefaultseppunct}\relax
\EndOfBibitem
\bibitem[Zhang \latin{et~al.}(2005)Zhang, Tan, Stormer, and Kim]{zhang2005experimental}
Zhang,~Y.; Tan,~Y.-W.; Stormer,~H.~L.; Kim,~P. Experimental observation of the quantum Hall effect and Berry's phase in graphene. \emph{nature} \textbf{2005}, \emph{438}, 201--204\relax
\mciteBstWouldAddEndPuncttrue
\mciteSetBstMidEndSepPunct{\mcitedefaultmidpunct}
{\mcitedefaultendpunct}{\mcitedefaultseppunct}\relax
\EndOfBibitem
\bibitem[Zhang \latin{et~al.}(2009)Zhang, Tang, Girit, Hao, Martin, Zettl, Crommie, Shen, and Wang]{zhang2009}
Zhang,~Y.; Tang,~T.-T.; Girit,~C.; Hao,~Z.; Martin,~M.~C.; Zettl,~A.; Crommie,~M.~F.; Shen,~Y.~R.; Wang,~F. Direct observation of a widely tunable bandgap in bilayer graphene. \emph{Nature} \textbf{2009}, \emph{459}, 820--823\relax
\mciteBstWouldAddEndPuncttrue
\mciteSetBstMidEndSepPunct{\mcitedefaultmidpunct}
{\mcitedefaultendpunct}{\mcitedefaultseppunct}\relax
\EndOfBibitem
\bibitem[Zibrov \latin{et~al.}(2017)Zibrov, Kometter, Zhou, Spanton, Taniguchi, Watanabe, Zaletel, and Young]{zibrov2017tunable}
Zibrov,~A.~A.; Kometter,~C.; Zhou,~H.; Spanton,~E.; Taniguchi,~T.; Watanabe,~K.; Zaletel,~M.; Young,~A. Tunable interacting composite fermion phases in a half-filled bilayer-graphene Landau level. \emph{Nature} \textbf{2017}, \emph{549}, 360--364\relax
\mciteBstWouldAddEndPuncttrue
\mciteSetBstMidEndSepPunct{\mcitedefaultmidpunct}
{\mcitedefaultendpunct}{\mcitedefaultseppunct}\relax
\EndOfBibitem
\bibitem[Qian \latin{et~al.}(2019)Qian, Lei, Wei, Zhang, Tang, Zhong, Zheng, and Chen]{qian2019}
Qian,~Q.; Lei,~J.; Wei,~J.; Zhang,~Z.; Tang,~G.; Zhong,~K.; Zheng,~Z.; Chen,~K.~J. 2D materials as semiconducting gate for field-effect transistors with inherent over-voltage protection and boosted ON-current. \emph{npj 2D Materials and Applications} \textbf{2019}, \emph{3}, 24\relax
\mciteBstWouldAddEndPuncttrue
\mciteSetBstMidEndSepPunct{\mcitedefaultmidpunct}
{\mcitedefaultendpunct}{\mcitedefaultseppunct}\relax
\EndOfBibitem
\bibitem[Gallagher \latin{et~al.}(2019)Gallagher, Yang, Lyu, Tian, Kou, Zhang, Watanabe, Taniguchi, and Wang]{gallagher2019}
Gallagher,~P.; Yang,~C.-S.; Lyu,~T.; Tian,~F.; Kou,~R.; Zhang,~H.; Watanabe,~K.; Taniguchi,~T.; Wang,~F. Quantum-critical conductivity of the Dirac fluid in graphene. \emph{Science} \textbf{2019}, \emph{364}, 158--162\relax
\mciteBstWouldAddEndPuncttrue
\mciteSetBstMidEndSepPunct{\mcitedefaultmidpunct}
{\mcitedefaultendpunct}{\mcitedefaultseppunct}\relax
\EndOfBibitem
\bibitem[Seo \latin{et~al.}(2024)Seo, Lu, Park, Yang, Xia, Ye, Yao, Han, Shi, Watanabe, \latin{et~al.} others]{seo2024on-chip}
Seo,~J.; Lu,~Z.; Park,~S.; Yang,~J.; Xia,~F.; Ye,~S.; Yao,~Y.; Han,~T.; Shi,~L.; Watanabe,~K.; others On-Chip Terahertz Spectroscopy for Dual-Gated van der Waals Heterostructures at Cryogenic Temperatures. \emph{Nano Letters} \textbf{2024}, \emph{24}, 15060--15067\relax
\mciteBstWouldAddEndPuncttrue
\mciteSetBstMidEndSepPunct{\mcitedefaultmidpunct}
{\mcitedefaultendpunct}{\mcitedefaultseppunct}\relax
\EndOfBibitem
\bibitem[Yan \latin{et~al.}(2012)Yan, Kim, Elle, Sushkov, Jenkins, Milchberg, Fuhrer, and Drew]{yan2012dual}
Yan,~J.; Kim,~M.~H.; Elle,~J.~A.; Sushkov,~A.~B.; Jenkins,~G.~S.; Milchberg,~H.~M.; Fuhrer,~M.~S.; Drew,~H. Dual-gated bilayer graphene hot-electron bolometer. \emph{Nature nanotechnology} \textbf{2012}, \emph{7}, 472--478\relax
\mciteBstWouldAddEndPuncttrue
\mciteSetBstMidEndSepPunct{\mcitedefaultmidpunct}
{\mcitedefaultendpunct}{\mcitedefaultseppunct}\relax
\EndOfBibitem
\bibitem[Yoshioka \latin{et~al.}(2022)Yoshioka, Wakamura, Hashisaka, Watanabe, Taniguchi, and Kumada]{yoshioka2022ultrafast}
Yoshioka,~K.; Wakamura,~T.; Hashisaka,~M.; Watanabe,~K.; Taniguchi,~T.; Kumada,~N. Ultrafast intrinsic optical-to-electrical conversion dynamics in a graphene photodetector. \emph{Nature Photonics} \textbf{2022}, \emph{16}, 718--723\relax
\mciteBstWouldAddEndPuncttrue
\mciteSetBstMidEndSepPunct{\mcitedefaultmidpunct}
{\mcitedefaultendpunct}{\mcitedefaultseppunct}\relax
\EndOfBibitem
\bibitem[Kim \latin{et~al.}(2013)Kim, Yan, Suess, Murphy, Fuhrer, and Drew]{kim2013photothermal}
Kim,~M.-H.; Yan,~J.; Suess,~R.; Murphy,~T.; Fuhrer,~M.; Drew,~H. Photothermal response in dual-gated bilayer graphene. \emph{Physical Review Letters} \textbf{2013}, \emph{110}, 247402\relax
\mciteBstWouldAddEndPuncttrue
\mciteSetBstMidEndSepPunct{\mcitedefaultmidpunct}
{\mcitedefaultendpunct}{\mcitedefaultseppunct}\relax
\EndOfBibitem
\bibitem[Gopalan \latin{et~al.}(2017)Gopalan, Janner, Nanot, Parret, Lundeberg, Koppens, Pruneri, \latin{et~al.} others]{gopalan2017mid}
Gopalan,~K.~K.; Janner,~D.~L.; Nanot,~S.; Parret,~R.; Lundeberg,~M.~B.; Koppens,~F.~H.; Pruneri,~V.; others Mid-Infrared Pyroresistive Graphene Detector on LiNbO3. \emph{Advanced Optical Materials} \textbf{2017}, \emph{5}, 1600723\relax
\mciteBstWouldAddEndPuncttrue
\mciteSetBstMidEndSepPunct{\mcitedefaultmidpunct}
{\mcitedefaultendpunct}{\mcitedefaultseppunct}\relax
\EndOfBibitem
\bibitem[Liu \latin{et~al.}(2016)Liu, Stradins, and Wei]{liu2016van}
Liu,~Y.; Stradins,~P.; Wei,~S.-H. Van der Waals metal-semiconductor junction: Weak Fermi level pinning enables effective tuning of Schottky barrier. \emph{Science advances} \textbf{2016}, \emph{2}, e1600069\relax
\mciteBstWouldAddEndPuncttrue
\mciteSetBstMidEndSepPunct{\mcitedefaultmidpunct}
{\mcitedefaultendpunct}{\mcitedefaultseppunct}\relax
\EndOfBibitem
\bibitem[Cloninger \latin{et~al.}(2024)Cloninger, Harris, Haley, Sterbentz, Taniguchi, Watanabe, and Island]{Cloninger_2024}
Cloninger,~J.~A.; Harris,~R.; Haley,~K.~L.; Sterbentz,~R.~M.; Taniguchi,~T.; Watanabe,~K.; Island,~J.~O. A back-to-back diode model applied to van der Waals Schottky diodes. \emph{Journal of Physics: Condensed Matter} \textbf{2024}, \emph{36}, 455301\relax
\mciteBstWouldAddEndPuncttrue
\mciteSetBstMidEndSepPunct{\mcitedefaultmidpunct}
{\mcitedefaultendpunct}{\mcitedefaultseppunct}\relax
\EndOfBibitem
\bibitem[Purdie \latin{et~al.}(2018)Purdie, Pugno, Taniguchi, Watanabe, Ferrari, and Lombardo]{purdie2018cleaning}
Purdie,~D.~G.; Pugno,~N.; Taniguchi,~T.; Watanabe,~K.; Ferrari,~A.; Lombardo,~A. Cleaning interfaces in layered materials heterostructures. \emph{Nature communications} \textbf{2018}, \emph{9}, 5387\relax
\mciteBstWouldAddEndPuncttrue
\mciteSetBstMidEndSepPunct{\mcitedefaultmidpunct}
{\mcitedefaultendpunct}{\mcitedefaultseppunct}\relax
\EndOfBibitem
\bibitem[Haley \latin{et~al.}(2021)Haley, Cloninger, Cerminara, Sterbentz, Taniguchi, Watanabe, and Island]{haley2021heated}
Haley,~K.~L.; Cloninger,~J.~A.; Cerminara,~K.; Sterbentz,~R.~M.; Taniguchi,~T.; Watanabe,~K.; Island,~J.~O. Heated assembly and transfer of van der Waals heterostructures with common nail polish. \emph{Nanomanufacturing} \textbf{2021}, \emph{1}, 49--56\relax
\mciteBstWouldAddEndPuncttrue
\mciteSetBstMidEndSepPunct{\mcitedefaultmidpunct}
{\mcitedefaultendpunct}{\mcitedefaultseppunct}\relax
\EndOfBibitem
\bibitem[Geim and Novoselov(2007)Geim, and Novoselov]{geim2007}
Geim,~A.; Novoselov,~K. The Rise of Graphene. \emph{Nature Mater} \textbf{2007}, \emph{6}, 183--191\relax
\mciteBstWouldAddEndPuncttrue
\mciteSetBstMidEndSepPunct{\mcitedefaultmidpunct}
{\mcitedefaultendpunct}{\mcitedefaultseppunct}\relax
\EndOfBibitem
\bibitem[Allen \latin{et~al.}(2010)Allen, Tung, and Kaner]{allen2010}
Allen,~M.~J.; Tung,~V.~C.; Kaner,~R.~B. Honeycomb Carbon: A Review of Graphene. \emph{Chemical Reviews} \textbf{2010}, \emph{110}, 132--145\relax
\mciteBstWouldAddEndPuncttrue
\mciteSetBstMidEndSepPunct{\mcitedefaultmidpunct}
{\mcitedefaultendpunct}{\mcitedefaultseppunct}\relax
\EndOfBibitem
\bibitem[Yan and Fuhrer(2010)Yan, and Fuhrer]{yan2010}
Yan,~J.; Fuhrer,~M.~S. Charge Transport in Dual Gated Bilayer Graphene with Corbino Geometry. \emph{Nano Letters} \textbf{2010}, \emph{10}, 4521--4525\relax
\mciteBstWouldAddEndPuncttrue
\mciteSetBstMidEndSepPunct{\mcitedefaultmidpunct}
{\mcitedefaultendpunct}{\mcitedefaultseppunct}\relax
\EndOfBibitem
\bibitem[Kuzmenko \latin{et~al.}(2009)Kuzmenko, van Heumen, van~der Marel, Lerch, Blake, Novoselov, and Geim]{kuzmenko2009}
Kuzmenko,~A.~B.; van Heumen,~E.; van~der Marel,~D.; Lerch,~P.; Blake,~P.; Novoselov,~K.~S.; Geim,~A.~K. Infrared spectroscopy of electronic bands in bilayer graphene. \emph{Phys. Rev. B} \textbf{2009}, \emph{79}, 115441\relax
\mciteBstWouldAddEndPuncttrue
\mciteSetBstMidEndSepPunct{\mcitedefaultmidpunct}
{\mcitedefaultendpunct}{\mcitedefaultseppunct}\relax
\EndOfBibitem
\bibitem[Efetov \latin{et~al.}(2011)Efetov, Maher, Glinskis, and Kim]{efetov2011}
Efetov,~D.~K.; Maher,~P.; Glinskis,~S.; Kim,~P. Multiband transport in bilayer graphene at high carrier densities. \emph{Phys. Rev. B} \textbf{2011}, \emph{84}, 161412\relax
\mciteBstWouldAddEndPuncttrue
\mciteSetBstMidEndSepPunct{\mcitedefaultmidpunct}
{\mcitedefaultendpunct}{\mcitedefaultseppunct}\relax
\EndOfBibitem
\bibitem[Gosling \latin{et~al.}(2021)Gosling, Makarovsky, Wang, Cottam, Greenaway, Patan{\`e}, Wildman, Tuck, Turyanska, and Fromhold]{gosling2021}
Gosling,~J.~H.; Makarovsky,~O.; Wang,~F.; Cottam,~N.~D.; Greenaway,~M.~T.; Patan{\`e},~A.; Wildman,~R.~D.; Tuck,~C.~J.; Turyanska,~L.; Fromhold,~T.~M. Universal mobility characteristics of graphene originating from charge scattering by ionised impurities. \emph{Communications Physics} \textbf{2021}, \emph{4}, 30\relax
\mciteBstWouldAddEndPuncttrue
\mciteSetBstMidEndSepPunct{\mcitedefaultmidpunct}
{\mcitedefaultendpunct}{\mcitedefaultseppunct}\relax
\EndOfBibitem
\bibitem[Cobaleda \latin{et~al.}(2014)Cobaleda, Pezzini, Diez, and Bellani]{cobaleda2014}
Cobaleda,~C.; Pezzini,~S.; Diez,~E.; Bellani,~V. Temperature- and density-dependent transport regimes in a $h$-BN/bilayer graphene/$h$-BN heterostructure. \emph{Phys. Rev. B} \textbf{2014}, \emph{89}, 121404\relax
\mciteBstWouldAddEndPuncttrue
\mciteSetBstMidEndSepPunct{\mcitedefaultmidpunct}
{\mcitedefaultendpunct}{\mcitedefaultseppunct}\relax
\EndOfBibitem
\bibitem[Tan \latin{et~al.}(2022)Tan, Adinehloo, Hone, and Perebeinos]{tan2022}
Tan,~C.; Adinehloo,~D.; Hone,~J.; Perebeinos,~V. Phonon-Limited Mobility in $h$-BN Encapsulated $AB$-Stacked Bilayer Graphene. \emph{Phys. Rev. Lett.} \textbf{2022}, \emph{128}, 206602\relax
\mciteBstWouldAddEndPuncttrue
\mciteSetBstMidEndSepPunct{\mcitedefaultmidpunct}
{\mcitedefaultendpunct}{\mcitedefaultseppunct}\relax
\EndOfBibitem
\bibitem[Shimazaki \latin{et~al.}(2015)Shimazaki, Yamamoto, Borzenets, Watanabe, Taniguchi, and Tarucha]{shimazaki2015}
Shimazaki,~Y.; Yamamoto,~M.; Borzenets,~I.~V.; Watanabe,~K.; Taniguchi,~T.; Tarucha,~S. Generation and detection of pure valley current by electrically induced Berry curvature in bilayer graphene. \emph{Nature Physics} \textbf{2015}, \emph{11}, 1032--1036\relax
\mciteBstWouldAddEndPuncttrue
\mciteSetBstMidEndSepPunct{\mcitedefaultmidpunct}
{\mcitedefaultendpunct}{\mcitedefaultseppunct}\relax
\EndOfBibitem
\bibitem[Henriksen and Eisenstein(2010)Henriksen, and Eisenstein]{henriksen2010}
Henriksen,~E.~A.; Eisenstein,~J.~P. Measurement of the electronic compressibility of bilayer graphene. \emph{Phys. Rev. B} \textbf{2010}, \emph{82}, 041412\relax
\mciteBstWouldAddEndPuncttrue
\mciteSetBstMidEndSepPunct{\mcitedefaultmidpunct}
{\mcitedefaultendpunct}{\mcitedefaultseppunct}\relax
\EndOfBibitem
\bibitem[Late \latin{et~al.}(2012)Late, Liu, Matte, Dravid, and Rao]{late2012}
Late,~D.~J.; Liu,~B.; Matte,~H. S. S.~R.; Dravid,~V.~P.; Rao,~C. N.~R. Hysteresis in Single-Layer MoS2 Field Effect Transistors. \emph{ACS Nano} \textbf{2012}, \emph{6}, 5635--5641\relax
\mciteBstWouldAddEndPuncttrue
\mciteSetBstMidEndSepPunct{\mcitedefaultmidpunct}
{\mcitedefaultendpunct}{\mcitedefaultseppunct}\relax
\EndOfBibitem
\bibitem[Cho \latin{et~al.}(2013)Cho, Park, Park, Jeong, Jang, Kim, Hong, Hong, and Lee]{cho2013}
Cho,~K.; Park,~W.; Park,~J.; Jeong,~H.; Jang,~J.; Kim,~T.-Y.; Hong,~W.-K.; Hong,~S.; Lee,~T. Electric Stress-Induced Threshold Voltage Instability of Multilayer MoS2 Field Effect Transistors. \emph{ACS Nano} \textbf{2013}, \emph{7}, 7751--7758\relax
\mciteBstWouldAddEndPuncttrue
\mciteSetBstMidEndSepPunct{\mcitedefaultmidpunct}
{\mcitedefaultendpunct}{\mcitedefaultseppunct}\relax
\EndOfBibitem
\bibitem[Lee \latin{et~al.}(2015)Lee, Cui, Kim, Arefe, Zhang, Lee, Ye, Watanabe, Taniguchi, Kim, and Hone]{lee2015}
Lee,~G.-H.; Cui,~X.; Kim,~Y.~D.; Arefe,~G.; Zhang,~X.; Lee,~C.-H.; Ye,~F.; Watanabe,~K.; Taniguchi,~T.; Kim,~P.; Hone,~J. Highly Stable, Dual-Gated MoS2 Transistors Encapsulated by Hexagonal Boron Nitride with Gate-Controllable Contact, Resistance, and Threshold Voltage. \emph{ACS Nano} \textbf{2015}, \emph{9}, 7019--7026\relax
\mciteBstWouldAddEndPuncttrue
\mciteSetBstMidEndSepPunct{\mcitedefaultmidpunct}
{\mcitedefaultendpunct}{\mcitedefaultseppunct}\relax
\EndOfBibitem
\bibitem[Jariwala \latin{et~al.}(2013)Jariwala, Sangwan, Late, Johns, Dravid, Marks, Lauhon, and Hersam]{jariwala2013}
Jariwala,~D.; Sangwan,~V.~K.; Late,~D.~J.; Johns,~J.~E.; Dravid,~V.~P.; Marks,~T.~J.; Lauhon,~L.~J.; Hersam,~M.~C. {Band-like transport in high mobility unencapsulated single-layer MoS2 transistors}. \emph{Applied Physics Letters} \textbf{2013}, \emph{102}, 173107\relax
\mciteBstWouldAddEndPuncttrue
\mciteSetBstMidEndSepPunct{\mcitedefaultmidpunct}
{\mcitedefaultendpunct}{\mcitedefaultseppunct}\relax
\EndOfBibitem
\bibitem[Bartolomeo \latin{et~al.}(2017)Bartolomeo, Genovese, Giubileo, Iemmo, Luongo, Foller, and Schleberger]{dibartolomeo2018}
Bartolomeo,~A.~D.; Genovese,~L.; Giubileo,~F.; Iemmo,~L.; Luongo,~G.; Foller,~T.; Schleberger,~M. Hysteresis in the transfer characteristics of MoS2 transistors. \emph{2D Materials} \textbf{2017}, \emph{5}, 015014\relax
\mciteBstWouldAddEndPuncttrue
\mciteSetBstMidEndSepPunct{\mcitedefaultmidpunct}
{\mcitedefaultendpunct}{\mcitedefaultseppunct}\relax
\EndOfBibitem
\bibitem[Lin \latin{et~al.}(2016)Lin, Kravchenko, Fowlkes, Li, Puretzky, Rouleau, Geohegan, and Xiao]{lin2016}
Lin,~M.-W.; Kravchenko,~I.~I.; Fowlkes,~J.; Li,~X.; Puretzky,~A.~A.; Rouleau,~C.~M.; Geohegan,~D.~B.; Xiao,~K. Thickness-dependent charge transport in few-layer MoS2 field-effect transistors. \emph{Nanotechnology} \textbf{2016}, \emph{27}, 165203\relax
\mciteBstWouldAddEndPuncttrue
\mciteSetBstMidEndSepPunct{\mcitedefaultmidpunct}
{\mcitedefaultendpunct}{\mcitedefaultseppunct}\relax
\EndOfBibitem
\bibitem[Radisavljevic and Kis(2013)Radisavljevic, and Kis]{radisavljevic2013}
Radisavljevic,~B.; Kis,~A. Mobility engineering and a metal--insulator transition in monolayer MoS2. \emph{Nature Materials} \textbf{2013}, \emph{12}, 815--820\relax
\mciteBstWouldAddEndPuncttrue
\mciteSetBstMidEndSepPunct{\mcitedefaultmidpunct}
{\mcitedefaultendpunct}{\mcitedefaultseppunct}\relax
\EndOfBibitem
\bibitem[Yang \latin{et~al.}(2020)Yang, Cai, Wu, and Fang]{yang2020}
Yang,~H.; Cai,~S.; Wu,~D.; Fang,~X. Humidity-Dependent Characteristics of Few-Layer MoS2 Field Effect Transistors. \emph{Advanced Electronic Materials} \textbf{2020}, \emph{6}, 2000659\relax
\mciteBstWouldAddEndPuncttrue
\mciteSetBstMidEndSepPunct{\mcitedefaultmidpunct}
{\mcitedefaultendpunct}{\mcitedefaultseppunct}\relax
\EndOfBibitem
\bibitem[Ji \latin{et~al.}(2019)Ji, Ghimire, Lee, Yi, Sakong, Gul, Yun, Jiang, Kim, Joo, Suh, and Lim]{ji2019}
Ji,~H.; Ghimire,~M.~K.; Lee,~G.; Yi,~H.; Sakong,~W.; Gul,~H.~Z.; Yun,~Y.; Jiang,~J.; Kim,~J.; Joo,~M.-K.; Suh,~D.; Lim,~S.~C. Temperature-Dependent Opacity of the Gate Field Inside MoS2 Field-Effect Transistors. \emph{ACS Applied Materials \& Interfaces} \textbf{2019}, \emph{11}, 29022--29028, PMID: 31313897\relax
\mciteBstWouldAddEndPuncttrue
\mciteSetBstMidEndSepPunct{\mcitedefaultmidpunct}
{\mcitedefaultendpunct}{\mcitedefaultseppunct}\relax
\EndOfBibitem
\bibitem[Trainer \latin{et~al.}(2022)Trainer, Nieminen, Bobba, Wang, Xi, Bansil, and Iavarone]{trainer2022visualization}
Trainer,~D.~J.; Nieminen,~J.; Bobba,~F.; Wang,~B.; Xi,~X.; Bansil,~A.; Iavarone,~M. Visualization of defect induced in-gap states in monolayer MoS2. \emph{npj 2D Materials and Applications} \textbf{2022}, \emph{6}, 13\relax
\mciteBstWouldAddEndPuncttrue
\mciteSetBstMidEndSepPunct{\mcitedefaultmidpunct}
{\mcitedefaultendpunct}{\mcitedefaultseppunct}\relax
\EndOfBibitem
\bibitem[Luryi(1988)]{luryi1988quantum}
Luryi,~S. Quantum capacitance devices. \emph{Applied Physics Letters} \textbf{1988}, \emph{52}, 501--503\relax
\mciteBstWouldAddEndPuncttrue
\mciteSetBstMidEndSepPunct{\mcitedefaultmidpunct}
{\mcitedefaultendpunct}{\mcitedefaultseppunct}\relax
\EndOfBibitem
\bibitem[Mak \latin{et~al.}(2010)Mak, Lee, Hone, Shan, and Heinz]{mak2010atomically}
Mak,~K.~F.; Lee,~C.; Hone,~J.; Shan,~J.; Heinz,~T.~F. Atomically thin MoS 2: a new direct-gap semiconductor. \emph{Physical review letters} \textbf{2010}, \emph{105}, 136805\relax
\mciteBstWouldAddEndPuncttrue
\mciteSetBstMidEndSepPunct{\mcitedefaultmidpunct}
{\mcitedefaultendpunct}{\mcitedefaultseppunct}\relax
\EndOfBibitem
\bibitem[Splendiani \latin{et~al.}(2010)Splendiani, Sun, Zhang, Li, Kim, Chim, Galli, and Wang]{splendiani2010emerging}
Splendiani,~A.; Sun,~L.; Zhang,~Y.; Li,~T.; Kim,~J.; Chim,~C.-Y.; Galli,~G.; Wang,~F. Emerging photoluminescence in monolayer MoS2. \emph{Nano letters} \textbf{2010}, \emph{10}, 1271--1275\relax
\mciteBstWouldAddEndPuncttrue
\mciteSetBstMidEndSepPunct{\mcitedefaultmidpunct}
{\mcitedefaultendpunct}{\mcitedefaultseppunct}\relax
\EndOfBibitem
\bibitem[Haley \latin{et~al.}(2021)Haley, Cloninger, Cerminara, Sterbentz, Taniguchi, Watanabe, and Island]{haley2021}
Haley,~K.~L.; Cloninger,~J.~A.; Cerminara,~K.; Sterbentz,~R.~M.; Taniguchi,~T.; Watanabe,~K.; Island,~J.~O. Heated Assembly and Transfer of Van der Waals Heterostructures with Common Nail Polish. \emph{Nanomanufacturing} \textbf{2021}, \emph{1}, 49--56\relax
\mciteBstWouldAddEndPuncttrue
\mciteSetBstMidEndSepPunct{\mcitedefaultmidpunct}
{\mcitedefaultendpunct}{\mcitedefaultseppunct}\relax
\EndOfBibitem
\bibitem[Golla \latin{et~al.}(2013)Golla, Chattrakun, Watanabe, Taniguchi, LeRoy, and Sandhu]{golla2013optical}
Golla,~D.; Chattrakun,~K.; Watanabe,~K.; Taniguchi,~T.; LeRoy,~B.~J.; Sandhu,~A. Optical thickness determination of hexagonal boron nitride flakes. \emph{Applied Physics Letters} \textbf{2013}, \emph{102}\relax
\mciteBstWouldAddEndPuncttrue
\mciteSetBstMidEndSepPunct{\mcitedefaultmidpunct}
{\mcitedefaultendpunct}{\mcitedefaultseppunct}\relax
\EndOfBibitem
\bibitem[Pierret \latin{et~al.}(2022)Pierret, Mele, Graef, Palomo, Taniguchi, Watanabe, Li, Toury, Journet, Steyer, Garnier, Loiseau, Berroir, Bocquillon, Fève, Voisin, Baudin, Rosticher, and Plaçais]{pierret2022}
Pierret,~A. \latin{et~al.}  Dielectric permittivity, conductivity and breakdown field of hexagonal boron nitride. \emph{Materials Research Express} \textbf{2022}, \emph{9}, 065901\relax
\mciteBstWouldAddEndPuncttrue
\mciteSetBstMidEndSepPunct{\mcitedefaultmidpunct}
{\mcitedefaultendpunct}{\mcitedefaultseppunct}\relax
\EndOfBibitem
\end{mcitethebibliography}


\providecommand{\noopsort}[1]{}\providecommand{\singleletter}[1]{#1}%
\providecommand{\latin}[1]{#1}
\makeatletter
\providecommand{\doi}
  {\begingroup\let\do\@makeother\dospecials
  \catcode`\{=1 \catcode`\}=2 \doi@aux}
\providecommand{\doi@aux}[1]{\endgroup\texttt{#1}}
\makeatother
\providecommand*\mcitethebibliography{\thebibliography}
\csname @ifundefined\endcsname{endmcitethebibliography}  {\let\endmcitethebibliography\endthebibliography}{}
\begin{mcitethebibliography}{4}
\providecommand*\natexlab[1]{#1}
\providecommand*\mciteSetBstSublistMode[1]{}
\providecommand*\mciteSetBstMaxWidthForm[2]{}
\providecommand*\mciteBstWouldAddEndPuncttrue
  {\def\EndOfBibitem{\unskip.}}
\providecommand*\mciteBstWouldAddEndPunctfalse
  {\let\EndOfBibitem\relax}
\providecommand*\mciteSetBstMidEndSepPunct[3]{}
\providecommand*\mciteSetBstSublistLabelBeginEnd[3]{}
\providecommand*\EndOfBibitem{}
\mciteSetBstSublistMode{f}
\mciteSetBstMaxWidthForm{subitem}{(\alph{mcitesubitemcount})}
\mciteSetBstSublistLabelBeginEnd
  {\mcitemaxwidthsubitemform\space}
  {\relax}
  {\relax}

\bibitem[Wang \latin{et~al.}(2010)Wang, Wu, Cong, Shang, and Yu]{wang2010}
Wang,~H.; Wu,~Y.; Cong,~C.; Shang,~J.; Yu,~T. Hysteresis of Electronic Transport in Graphene Transistors. \emph{ACS Nano} \textbf{2010}, \emph{4}, 7221--7228, PMID: 21047068\relax
\mciteBstWouldAddEndPuncttrue
\mciteSetBstMidEndSepPunct{\mcitedefaultmidpunct}
{\mcitedefaultendpunct}{\mcitedefaultseppunct}\relax
\EndOfBibitem
\bibitem[Zhang \latin{et~al.}(2009)Zhang, Tang, Girit, Hao, Martin, Zettl, Crommie, Shen, and Wang]{zhang2009}
Zhang,~Y.; Tang,~T.-T.; Girit,~C.; Hao,~Z.; Martin,~M.~C.; Zettl,~A.; Crommie,~M.~F.; Shen,~Y.~R.; Wang,~F. Direct observation of a widely tunable bandgap in bilayer graphene. \emph{Nature} \textbf{2009}, \emph{459}, 820--823\relax
\mciteBstWouldAddEndPuncttrue
\mciteSetBstMidEndSepPunct{\mcitedefaultmidpunct}
{\mcitedefaultendpunct}{\mcitedefaultseppunct}\relax
\EndOfBibitem
\bibitem[Qian \latin{et~al.}(2019)Qian, Lei, Wei, Zhang, Tang, Zhong, Zheng, and Chen]{qian2019}
Qian,~Q.; Lei,~J.; Wei,~J.; Zhang,~Z.; Tang,~G.; Zhong,~K.; Zheng,~Z.; Chen,~K.~J. 2D materials as semiconducting gate for field-effect transistors with inherent over-voltage protection and boosted ON-current. \emph{npj 2D Materials and Applications} \textbf{2019}, \emph{3}, 24\relax
\mciteBstWouldAddEndPuncttrue
\mciteSetBstMidEndSepPunct{\mcitedefaultmidpunct}
{\mcitedefaultendpunct}{\mcitedefaultseppunct}\relax
\EndOfBibitem
\end{mcitethebibliography}

\newpage

\begin{figure}
    \centering
    \includegraphics{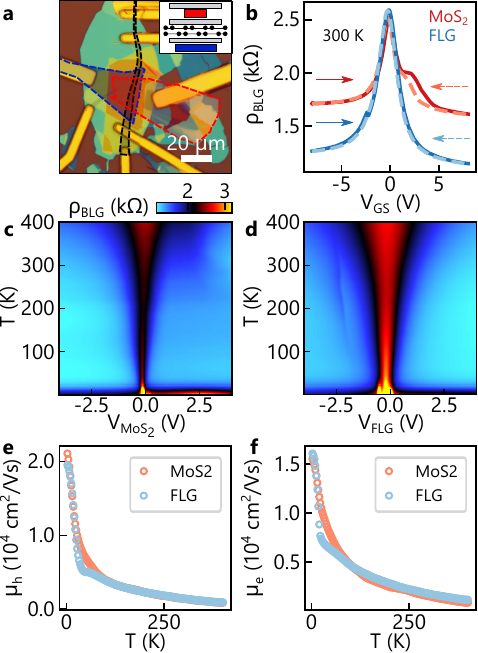}
    \caption{Graphite and MoS$_2$ gate comparison. \textbf{(a)} Optical microscope image of device. Outlined are the MoS$_2$ top gate (red), bilayer graphene (black), and graphite bottom gate (blue). Inset: Stacking order of the dual-gated graphene bilayer device with colors corresponding to outlines in main panel. Gray layers represent dielectric layers of h-BN. \textbf{(b)} Four-probe resistivity of the bilayer graphene as a function of gate-source voltage for MoS$_2$ gate (red) and FLG gate (blue) at 300 K. Solid lines represent forward sweeps, while dashed lines represent backward sweeps. \textbf{(c-d)} Resistivity of the bilayer graphene mapped as a function of temperature and MoS$_2$ gate (c) or FLG gate (d). Color bar above (c) applies to (d) as well. For each gate sweep in (b-d), the inactive gate was held at zero volts. \textbf{(e-f)} Field effect mobility of the holes (e) and electrons (f) in bilayer graphene as a function of temperature. The legend indicates whether the value is extracted from the MoS$_2$ data (c) or the FLG data (d).}
    \label{fig:1}
\end{figure}

\begin{figure*}
    \centering
    \includegraphics{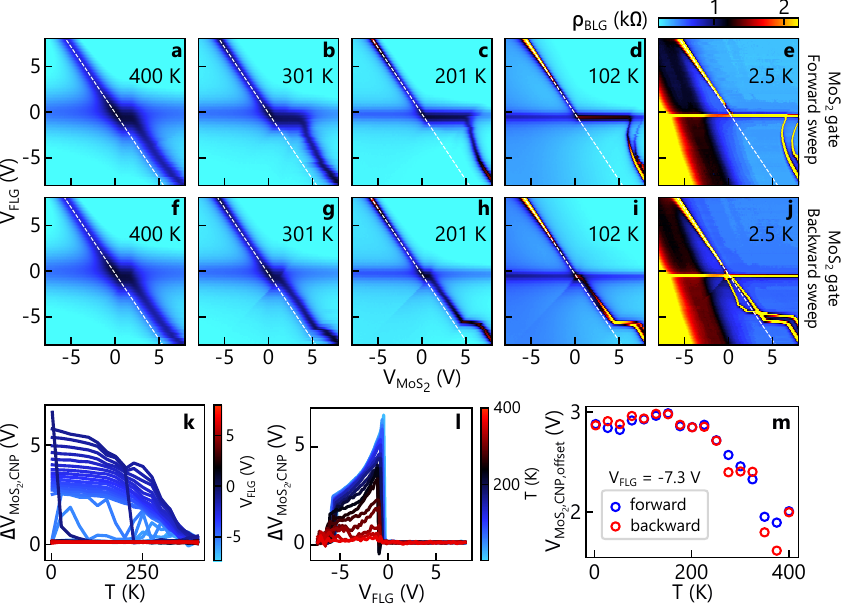}
    \caption{Dual-gated bilayer graphene sweeps. \textbf{(a-e)} BLG resistivity as a function of MoS$_2$ and FLG gate voltages. MoS$_2$ gate was swept along the fast axis forward (negative to positive voltage). The panels show the CNP evolution as a function of temperature, from 400 K (a) to 2.5 K (e). White dashed lines are linear fits to the CNP representing the expected gate voltages the CNP should appear. \textbf{(f-j)} Same as panels (a-e), but during the backward sweep (positive to negative voltage) of the MoS$_2$ gate. Color bar above (e) applies to (a-j). \textbf{(k-l)} Hysteresis of BLG CNP appearance in MoS$_2$ gate voltage as a function of temperature and the graphite gate voltage: (k) shows the temperature dependence, while (l) shows the graphite gate dependence. \textbf{(m)} CNP offset from expected MoS$_2$ voltage at V$_{FLG}$ = -7.3 V as a function of temperature for the MoS$_2$ forward sweep (blue) and backward sweep (red).}
    \label{fig:2}
\end{figure*}

\begin{figure}
    \centering
    \includegraphics{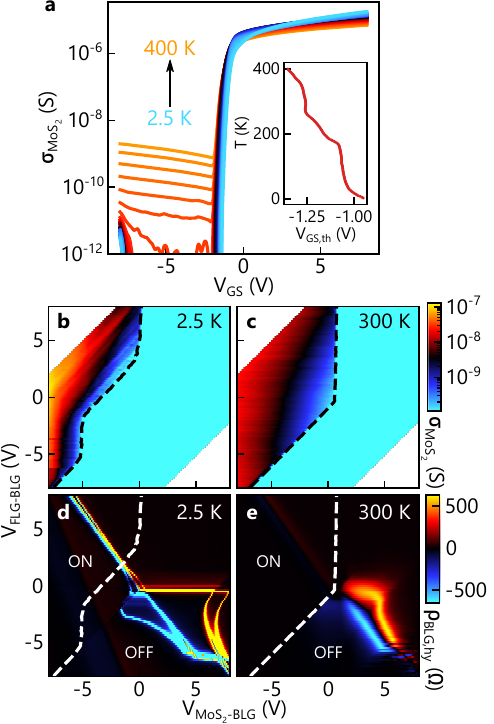}
    \caption{MoS$_2$ gate threshold and gating effectiveness. \textbf{(a)} MoS$_2$ conductivity as a function of gate-source voltage applied to electrically-connected BLG and FLG layers acting as a single back gate. Scans were taken from 2.5 K to 400 K, in steps of 7.5 K. Inset: Extracted threshold voltage as a function of temperature. \textbf{(b-c)} MoS$_2$ conductivity as a function of gate voltage relative to the bilayer graphene at 2.5 K (b) and 300 K (c). Dashed line indicates the MoS$_2$ threshold voltage. \textbf{(d-e)} Hysteresis of BLG resistivity (forward sweep minus backward sweep) as a function of gate voltage relative to BLG at 2.5 K (d) and 300 K (e). Dashed lines are the same as from panels (b-c).}
    \label{fig:3}
\end{figure}

\begin{figure}
    \centering
    \includegraphics{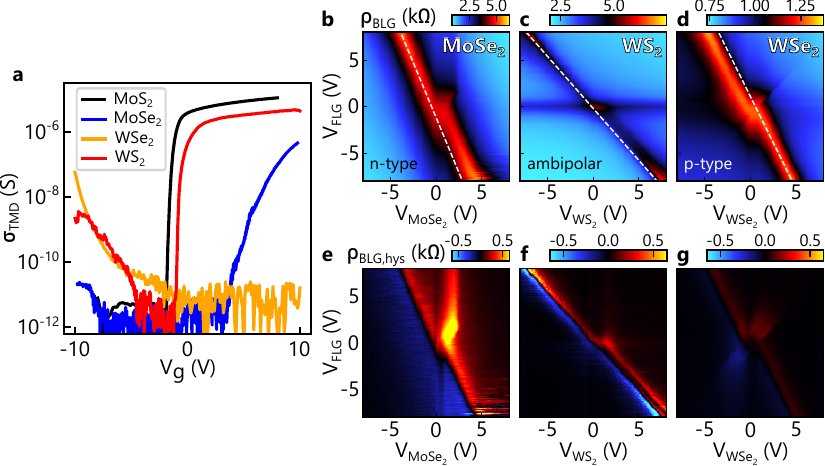}
    \caption{Gating bilayer graphene with MoSe$_2$, WS$_2$, and WSe$_2$ at room temperature. \textbf{(a)} Two terminal conductivity at room temperature for all four materials used as a semiconducting gate, as a function of gate-source voltage applied to electrically-connected BLG and FLG layers acting as a single back gate. \textbf{(b-d)} BLG resistivity as a function of MoSe$_2$(b), WS$_2$(c), and WSe$_2$(d), and FLG gate voltages. \textbf{(e-g)} Hysteresis of BLG resistivity (forward sweep minus backward sweep) as a function of MoSe$_2$(e), WS$_2$(f), and WSe$_2$(g) gate voltage relative to BLG at room temperature. For all 2D gate sweeps, the TMD gate voltage was the fast sweep axis. }
    \label{fig:4}
\end{figure}

\end{document}


\maketitle

\section{Supplementary Note 1: Device summary}\label{device_fab}


\begin{figure}[!]
    \centering
    \includegraphics[width=5 in]{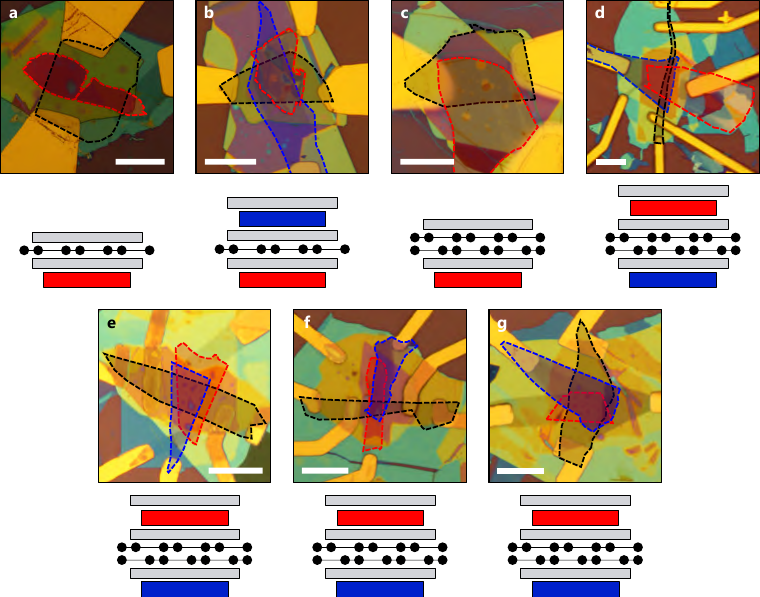}
    \caption{Optical microscope images and stacking profile of all devices reported in this manuscript. \textbf{(a)} TMDG008, \textbf{(b)} TMDG009, \textbf{(c)} TMDG017, \textbf{(d)} TMDG022, \textbf{(e)} TMDG023, \textbf{(f)} TMDG024, and \textbf{(g)} TMDG026. Black outline in microscope image corresponds to the graphene layer, red outline to the TMD gate layer, and blue outline to the few-layer graphite gate layer. Layers in the stacking profile correspond to outlines of like color, with gray layers representing h-BN dielectrics. Details of each device are described in Table S1.}
    \label{fig:S0}
\end{figure}

\begin{table}[!ht]
    \small
    \centering
    \caption{Summary of all devices measured. Graphene Thickness refers to if the device has a monolayer or bilayer graphene channel. TMD Material identifies the material used for the semiconducting gate. TMD Thickness is the thickness of the semiconducting gate. (l$_G$ $\times$ w$_G$) is the length and width of the graphene or bilayer graphene channel. (l$_{TMD}$ $\times$ w$_{TMD}$) is the length and width of the semiconducting gate. N/A refers to devices that did not have a second contact on the TMD layer for independent measurement.  }
    \begin{tabular}{|c|c|c|c|c|c|}
    \hline
        \textbf{Name} & \textbf{\begin{tabular}{@{}c@{}}Graphene \\ Thickness\end{tabular}} & \textbf{\begin{tabular}{@{}c@{}}TMD \\ Material\end{tabular}} & \textbf{\begin{tabular}{@{}c@{}}TMD Thickness \\ (layers)\end{tabular}} & \textbf{(l$_G$ $\times$ w$_G$) (\textmu m)} & \textbf{\begin{tabular}{@{}c@{}}(l$_{TMD}$ $\times$ \\ w$_{TMD}$) (\textmu m)\end{tabular}} \\ \hline
        \textbf{TMDG008} & monolayer & MoSe$_2$ & bulk & (10.7, 17.0) & N/A \\ \hline
        \textbf{TMDG009} & monolayer & MoS$_2$ & bulk & (15.7, 17.5) & N/A \\ \hline
        \textbf{TMDG017} & bilayer & MoSe$_2$ & bulk & (26.5, 17.1) & N/A \\ \hline
        \textbf{TMDG022} & bilayer & MoS$_2$ & 2 & (32.0, 3.5) & (20.0, 20.0) \\ \hline
        \textbf{TMDG023} & bilayer & MoSe$_2$ & 6 & (22.6, 11.7) & (22.8, 13.4) \\ \hline
        \textbf{TMDG024} & bilayer & WSe$_2$ & 3 & (18.5, 7.0) & (24.8, 5.6) \\ \hline
        \textbf{TMDG026} & bilayer & WS$_2$ & bulk & (37.0, 15.7) & (15.7, 11.9) \\ \hline
    \end{tabular}
\end{table}


\newpage

\section{Supplementary Note 2: Dual-gated monolayer graphene device with an MoS$_2$ gate}

\begin{figure}[!]
    \centering
    \includegraphics[width=5 in]{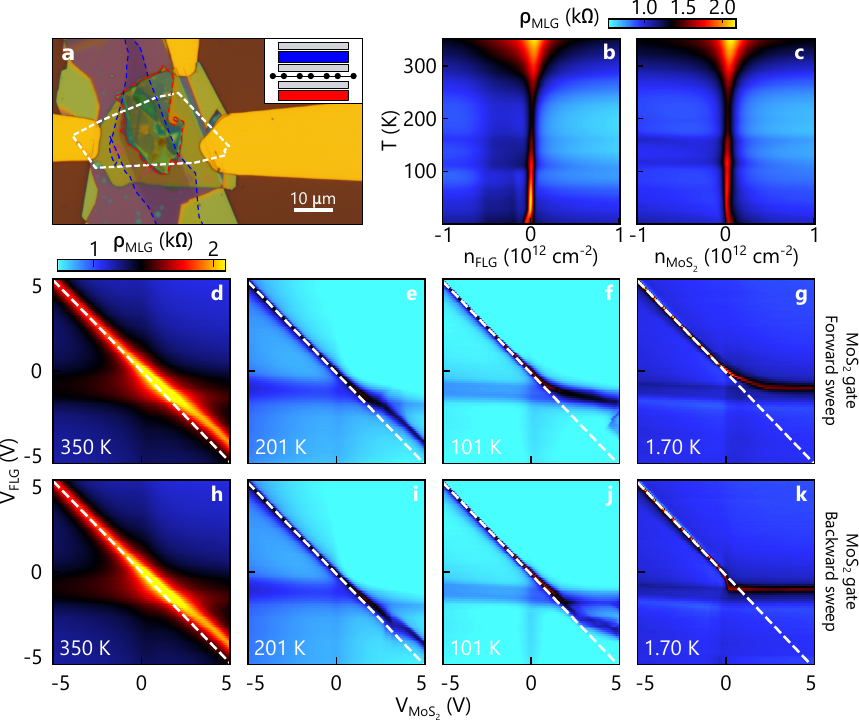}
    \caption{Characterization of a second MoS$_2$-gated device (TMDG009) with monolayer graphene instead of bilayer. \textbf{(a)} Optical image of device. Outlined are the FLG top gate (blue), monolayer graphene (MLG) sample (white), and multi-layer MoS$_2$ gate (red). Inset: stacking order of device, showing the FLG gate (blue), MLG layer (black ball and stick), and MoS$_2$ gate (red), all separated by h-BN layers (gray). \textbf{(b-c)} Resistivity of the MLG layer as a function of temperature and the carrier density induced by the FLG gate (b) and the MoS$_2$ gate (c). \textbf{(d-g)} MLG resistivity as a function of the MoS$_2$ (forward sweep) and FLG gates, from temperatures of 350 K (d) down to 1.7 K (g). Superimposed dashed lines indicate linear trend in CNP where the MoS$_2$ gate is negative. A deviation from that trend in the positive MoS$_2$ gate region matches the results of the main text device. \textbf{(h-k)} Same as (d-g) but during the backward sweep of the MoS$_2$ gate. Color scale above panel (d) applies to (d-k).}
    \label{fig:S7}
\end{figure}
\newpage
\section{Supplementary Note 3: Bilayer graphene device with alternative fast sweep axis}

\begin{figure}
    \centering
    \includegraphics{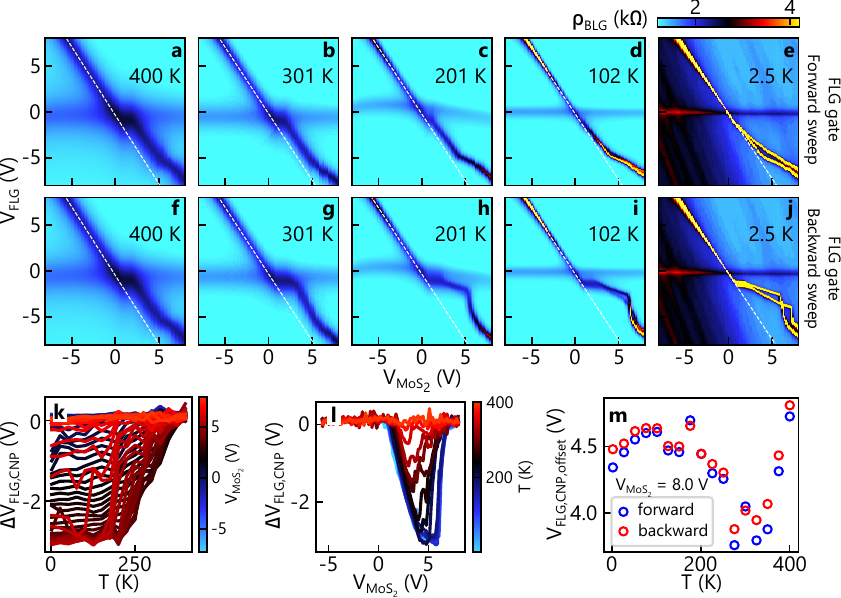}
    \caption{Similar to Figure \ref{fig:2} in the main text, but with the FLG gate as the fast gate sweep axis. \textbf{(a-j)} Many of the same trends are present as when the MoS$_2$ gate was swept fast: Higher temperature sweeps ((a),(f)) show a slight deviation in the CNP gate dependence from linear near zero gate voltages, while lower temperatures ((e),(j)) show a more significant deviation. All sweeps show a linear CNP gate dependence with negative MoS$_2$ and positive FLG gate voltages applied. \textbf{(k)} The CNP hysteresis, this time with regards to the FLG gate voltage, shows a similar trend of low hysteresis at high temperatures, with gradually worsening hysteresis as temperature decreases. \textbf{(l)} When plotted against the MoS$_2$ gate voltage, at significantly high positive voltage, hysteresis again drops. \textbf{(m)} CNP offset from expected FLG voltage at $V_{MoS_2}$ = 8 V as a function of temperature for the FLG forward sweep (blue) and backward sweep (red).}
    \label{fig:S1}
\end{figure}
\newpage


\section{Supplementary Note 4: Bilayer graphene device leakage currents}

\begin{figure}
    \centering
    \includegraphics{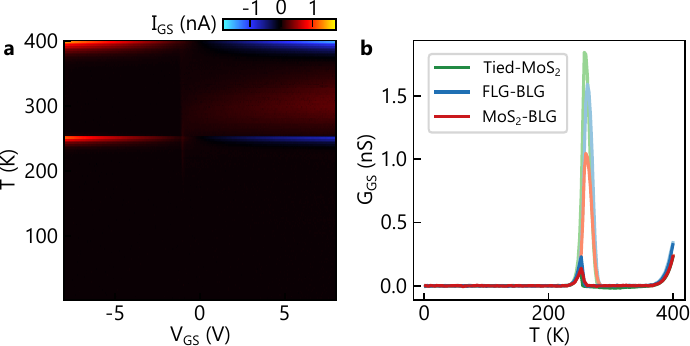}
    \caption{Leakage current through dielectric between gates. \textbf{(a)} Sample leakage current check from the MoS$_2$ gate to the BLG and FLG gate tied together, as a function of gate voltage and temperature. The increase in leakage at high temperature is expected as higher thermal energy encourages transport through the insulating h-BN layers. Unexpected, however, is the increase around 260 K. \textbf{(b)} Differential conductance through dielectric layers between gates as a function of temperature for different pairings of layers (green: BLG and FLG to MoS$_2$; blue: FLG to BLG; red: MoS$_2$ to BLG). The lines with darker colors were taken while sweeping the temperature up, while the lighter colors were taken during the down sweep. The anomaly around 260 K is more prevalent during the down sweep. We conjecture the source may be due to water accumulating on the surface of the device, potentially shorting electrodes through the poor conductivity of the water as it freezes.}
    \label{fig:S4}
\end{figure}
\newpage
\section{Supplementary Note 5: Bilayer graphene device hysteresis characterization}

\begin{figure}
    \centering
    \includegraphics[width=4 in]{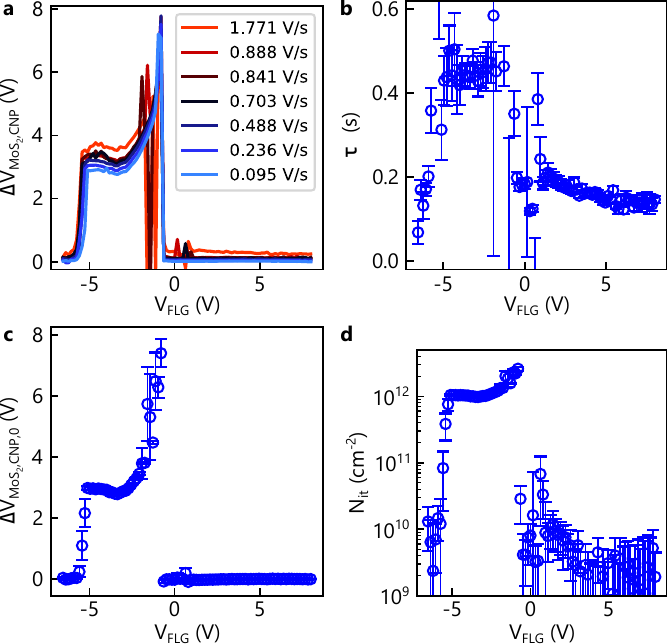}
    \caption{Hysteresis dependence of gate voltage sweep rate. \textbf{(a)} Hysteresis of CNP appearance in MoS$_2$ gate voltage (forward sweep minus backward sweep) as a function of the FLG gate voltage (similar to Figure \ref{fig:2}(l)) for different MoS$_2$ gate sweep rates. All sweeps performed at 2.5 K. \textbf{(b-c)} Fit parameters $\tau$ (b) and $\Delta V_{MoS_2,\textrm{CNP},0}$ (c) extracted from (a) by fitting the hysteresis vs sweep rate to a line ($\Delta V_{MoS_2,\textrm{CNP}} = \tau\dot{V}_{MoS_2} + \Delta V_{MoS_2,\textrm{CNP},0}$, where $\dot{V}_{MoS_2}$ is the sweep rate). The parameter $\tau$ gives us a temporal term related to how much the hysteresis is affected by sweep rate. The on-state (positive $V_{FLG}$) has a time of about 0.15 s, while the off-state is roughly at 0.45 s. The parameter $\Delta V_{MoS_2,\textrm{CNP},0}$ gives us information about the intrinsic hysteresis of the CNP if we were able to sweep the gate infinitely slowly (near 0 V/s). We would expect this value to be zero if the hysteresis was solely due to the gate being swept faster than the time constant the gates can charge at, as we see in the on-state (and slightly past the off-state). Nonzero values indicate some internal mechanism (such as charge trapping) leading to the MoS$_2$ locking into a certain charge state and lacking the ability for charge transfer to occur. \textbf{(d)} Trap state density extracted from (c) via $N_{it}=C'_{MoS_2\mhy BLG}\Delta V_{MoS_2,\textrm{CNP},0}/(2e)$\cite{wang2010}, where $C'_{MoS_2\mhy BLG}$ is the capacitance per area between the MoS$_2$ and BLG layers.}
    \label{fig:S5}
\end{figure}
\newpage
\section{Supplementary Note 6: Bilayer graphene device bandgap}

\begin{figure}
    \centering
    \includegraphics{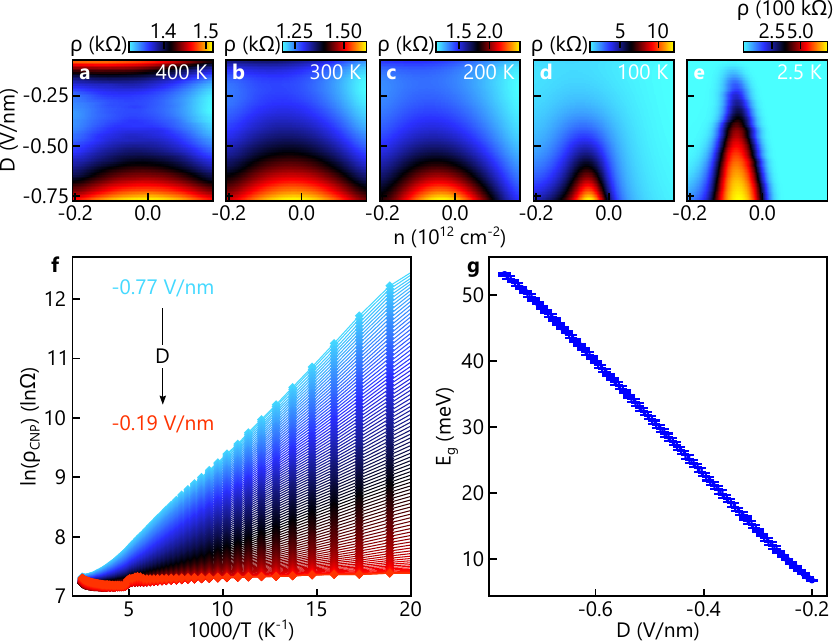}
    \caption{Bilayer graphene band gap opening. \textbf{(a-e)} Resistivity of BLG as a function of carrier density $n$ and displacement field $D$, for temperatures from 400 K (a) down to 2.5 K (e). The displacement field (which would more accurately be described as a screened electric field) is calculated by $D = \frac{1}{2\epsilon_0}\left(C'_{MoS_2}V_{MoS_2} - C'_{FLG}V_{FLG}\right)$\cite{zhang2009}. Negative displacement fields correspond to the on-state of the MoS$_2$ gate. \textbf{(f)} Arrhenius-like plots of the resistivity at CNP vs temperature, plotted for each displacement field from -0.77 V/nm to -0.19 V/nm. Clear linear trends are present in this temperature range (50 K to 400 K), allowing us to fit to the Arrhenius model: $\ln{\rho_{\textrm{CNP}}}=\frac{E_g}{2k_B}T^{-1}+\ln{\rho_{0}}$, where $E_g$ is the band gap opening and $k_B$ is the Boltzmann constant. \textbf{(g)} Band gap opening in BLG as a function of applied displacement field.}
    \label{fig:S6}
\end{figure}
\newpage

\section{Supplementary Note 7: One-dimensional voltage clamp model}
\begin{figure}
    \centering
    \includegraphics{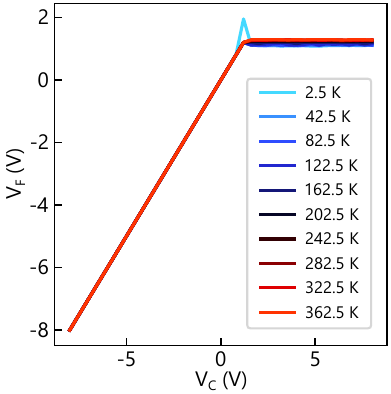}
    \caption{Simulation of the voltage on MoS$_2$ ($V_F$) a distance 10 \textmu m from the contact as a function of the applied contact voltage ($V_C$).}
    \label{fig:S9}
\end{figure}

To support the idea of the voltage on the MoS$_2$ gate clamping beyond the threshold voltage, we follow a method laid forth by Qian, et al.,\cite{qian2019} for a 1D gate model. In this model, we simplify the stack to a single an MoS$_2$ flake separated from the BLG-FLG bottom gate by an h-BN dielectric, whose length extends in the $x$ direction. The MoS$_2$ is contacted by a metal electrode at $x=0$. The potential across the flake $V_G(x)$ establishes an equilibrium under the following two conditions:
\begin{equation}\label{currdx}
    \frac{dI_G(x)}{dx} = J_{BN}(V_G(x))
\end{equation}
\begin{equation}\label{voltdx}
    \frac{dV_G(x)}{dx} = \frac{I_G(x)}{G_{MoS_2}(V_G(x))}
\end{equation}
Equation \ref{currdx} states the rate of change for the linear leakage current $I_G$ (normalized by the channel width) with respect to position is equal to the leakage current density $J_{BN}$. Equation \ref{voltdx} states the rate of voltage change with respect to position is equal to the ohmic voltage change due to a linear leakage current passing through MoS$_2$ with a conductance $G_{MoS_2}$ at that gate potential.
Empirical fits were estimated to model $J_{BN}$ and $G_{MoS_2}$ for ease of calculation. For $J_{BN}$, linear relations were sufficient to fit leakage current measurements between the BLG and FLG layers (see Figure \ref{fig:S4} (a) for example leakage data). For $G_{MoS_2}$, we adapt Qian et al.'s expression for a metal-oxide-semiconductor structure and modify it to account for a transition from a linear to a saturated regime we observed in our MoS$_2$ transconductance curves:
\begin{equation}
    G_{MoS_2}(V_G) = \mu_{MoS_2}C'_{BN}\frac{W_{MoS_2}}{L_{MoS_2}}\frac{nk_BT}{q}\ln{\left[1 + e^{\frac{q(V_G-V_{th})}{nk_BT}}\right]}\left\{\frac{r+1}{2}+\frac{r-1}{\pi} \arctan{\left[m(V_G-V_S)\right]}\right\}
\end{equation}
where $\mu_{MoS_2}$ is the mobility of MoS$_2$, $C'_{BN}$ is the capacitance density between the layers, $n$ is a fitting parameter describing the subthreshold-linear transition, $W_{MoS_2}$ and $L_{MoS_2}$ are the width and length of the MoS$_2$ gate, $k_B$ is the Boltzmann constant, $T$ is temperature, $q$ is the elementary charge, $V_{th}$ is the MoS$_2$ threshold voltage, $r$ is a fit parameter that modifies the linear slope when in the saturated regime, $m$ is a fitting parameter describing the linear-saturated transition, and $V_S$ is the gate voltage at which the linear-saturated transition occurs. Mathematically, the natural log term transitions from zero in the subthreshold regime to the linear regime described by $G_{lin}(V_G)=\mu_{MoS_2}C'_{BN}(V_G-V_{th})\frac{W_{MoS_2}}{L_{MoS_2}}$. The arctangent term then transitions to the saturated regime where the linear slope $\mu_{MoS_2}C'_{BN}\frac{W_{MoS_2}}{L_{MoS_2}}$ is multiplied by the factor $r\in[0,1]$.
Equations \ref{currdx} and \ref{voltdx} can be combined into an integral expression via separation of variables:
\begin{equation}\label{zero1}
    I_G(x) = \sqrt{\int_{V_G(0)}^{V_G(x)} 2J_{BN}(V_G)G(V_G)dV_G}
\end{equation}
Applying the same treatment to Equation \ref{voltdx} alone:
\begin{equation}\label{zero2}
    x = \frac{1}{I_G(x)}\int_{V_G(0)}^{V_G(x)}G_{MoS_2}(V_G)dV_G
\end{equation}
Equations \ref{zero1} and \ref{zero2} can now be solved numerically by assuming a contact voltage $V(0) = V_C$ and solving for the floating voltage $V(x) = V_F$ and current $I(x)$. The results of this model are shown in Figure \ref{fig:S9}, where the floating voltage is determined at a fixed distance $x = 10$ \textmu m from the contact for a sweep of contact voltages at various temperatures. The simulation shows a clamping of the voltage beyond a threshold of about 1.2 V, matching the behavior described in the main text.

\newpage

\newpage
\section{Supplementary Note 8: Monolayer and bilayer graphene devices with a MoSe$_2$ gate}

\begin{figure}
    \centering
    \includegraphics[width=3.6 in]{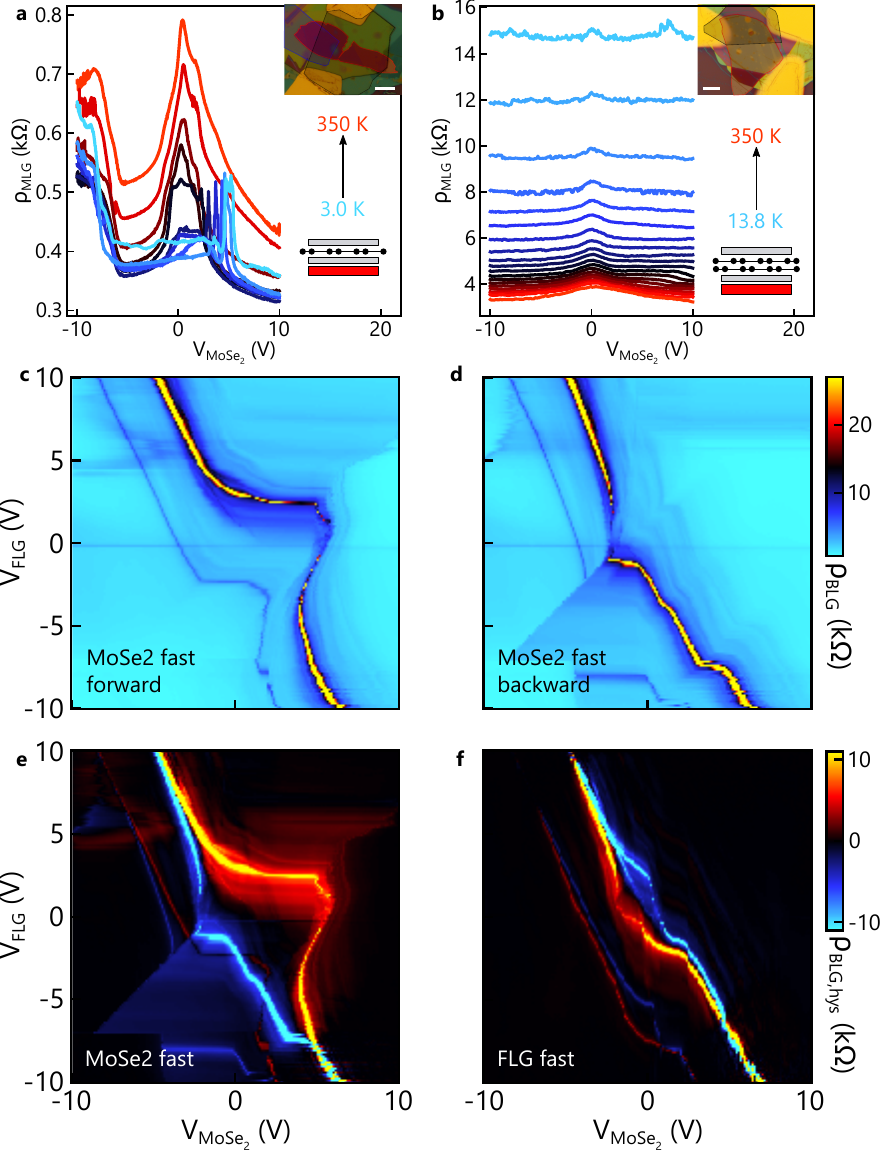}
    \caption{Measurements of additional graphene devices with a MoSe$_2$ gate at various temperatures. \textbf{(a)} MoSe$_2$-gated device (TMDG008). The flake is cracked in he middle of the graphene channel. Inset upper: optical image. Inset lower: stacking schematic (blue: FLG; red: MoSe$_2$). \textbf{(b)} MoSe$_2$-gated device (TMDG017). FLG gate was held at 0 V during sweep of MoSe2 gate. Inset upper: optical image. Inset lower: stacking schematic (red: MoSe$_2$). \textbf{(c-f)} 2D gate sweeps of device TMDG023 taken at low temperature (2 K). Gate sweep taken with the MoSe$_2$ gate voltage as the fast axis; (c) is the bilayer graphene resistivity during the forward sweep, (d) is during the backward sweep, and (e) is the difference between forward resistivity and backward resistivity, similar to Figure 3(d,e) in the main text. (f) Hysteresis of forward sweep resistivity minus the backward sweep, with the few-layer graphite gate voltage as the fast axis.}
    \label{fig:S8}
    %
    %
    %
    %
\end{figure}

\newpage

\bibliography{main_text}